\documentclass[preprint,aps,pra,authoryear]{revtex4}
\usepackage{tabularx}
\usepackage{dcolumn}
\usepackage{graphics}
\usepackage{bm}
\usepackage{graphicx}
\usepackage[hypertex]{hyperref}
\usepackage{tabularx}
\usepackage{amsmath}
\usepackage{amssymb}
\usepackage{longtable}
\usepackage{verbatim}
\usepackage{float} 
\usepackage{fancyhdr} 
\usepackage{epstopdf}

\begin{document}
\pagestyle{fancy}
\rhead{\thepage}
\chead{}
\lhead{}
\cfoot{}
\rfoot{}
\lfoot{}

\title{Quarkonium in a spin dependent quark anti-quark potential at finite temperature}
\author{Zunara Hussain, Faisal Akram, Bilal Masud}\affiliation{Centre For High Energy Physics, Punjab
University, Lahore(54590), Pakistan. }

\begin{abstract}
We study the states of $c\bar{c}$ and $b\bar{b}$ system in a non-relativistic a spin dependent potential model at finite temperature. The potential incorporate color screening effect through a parameter $\mu$. The parameter $\mu$ along with strong coupling constant $\alpha_s$ and string tension $\sigma$ are taken temperature dependent and their values are fitted to the Lattice data of quark anti-quark potential available at different values of temperature. The applied potential include spin-spin, spin-orbital, tensor interactions which allow us to study spin singlet and triplet states of quarkonia. By solving non-relativistic Schr{\"o}dinger equation, we find the spectrum of quarkonium states at different temperature and also determine the critical temperature at which each state is dissolved. We find the critical temperature is decreased with orbital and radial excitations, however, for spin singlet and triplet states it remains same. 
\end{abstract}
\maketitle

\section{Introduction}
Quantum Chromodynamics (QCD) describes strongly interacting particles (quarks and gluons). According to QCD, when temperature of the hadronic matter increases beyond a certain value then hadronic matter will undergo a phase transition to a new state of matter called Quark Gluon Plasma~\cite{pal}. It is a deconfined state of quarks and gluons. This deconfinement is due to the color charges producing color screening resulting in a reduced binding potential between quarks and anti-quarks~\cite{Rosi}. This screening force is similar to the QED Debye screening that affects electric charges in QED. The ground state of heavy quarkonia (\emph{b}$\bar{b}$  and \emph{c}$\bar{c}$ ) has greater binding energy than that of lighter mesons and nucleons so they can survive in QGP unless the temperature of the medium is high enough to break the binding between quarks and antiquarks~\cite{Helmut}. So one of the tool to investigate the QGP is to analyze the properties of heavy quarkonia at finite temperature.
\paragraph*{}
 Matasui and H. Satz~\cite{Matsui}\cite{satz} considered suppression of $J/\psi$ as a signal of QGP formation. Zahed and Shuryak~\cite{shuryak}\cite{shuryak1}\cite{zahed} proposed that ground state of heavy quarkonia are not effected in Quark Gluon Plasma unless temperature of the medium increased further.
  They approximate the binding energy of charmonium ground state
  and predict that $J\psi$ does not dissociate
  unless the temperature reaches 2.7 $T_c$~\cite{zahed}.
   Blaizot et al.~\cite{blaizot} statistically investigate the confinement and deconfinement of heavy quarkonia in QGP
   They observed that if heavy quarks are found in the system then the formation of bound state
   occurs,
   while  dissociation occurs
   due to potential screening in the plasma. Kisslinger~\cite{kisslinger} 
    presents a detailed review on QCD, QGP and heavy quarkonium suppression. Grandchamp et al.~\cite{grandchamp} discussed the charmonium production in heavy-ion collisions. They used lattice gauge theory results to model the effect of medium modifications in the charm quark and checked
     the dissociation and recreation of charmonium throughout the reaction. Park~\cite{park} observed
      the mechanism of heavy quaronium dissociation. He
      has pointed out that in the deconfined medium, by increasing the chemical potential dissociation length decreased while it has the opposite trend in the confined phase. By using maximum entropy and spectral function analysis, Petreczky et al.~\cite{datta} and Asakawa et al.~\cite{asakawa} observed the widths and masses of heavy quark bound states in the frame work of quenched LQCD. They observed that dissociation temperature of $J\psi$ is greater than 1.6 times the critical temperature of QGP.
 \paragraph*{}
  Bo.Liu and Yu.Bing Dong~\cite{b.liu} have used different non-relativistic~\emph{q}$\bar{q}$ potentials as a function of medium temperature for studying the binding radii and masses of heavy quarkonia bound states.
   By using Debye color screening effect they estimated the critical value of screening length and critical temperature of the medium. F. Karsch also used potential model as a function of temperature dependent parameter $\mu$ in Schr{\"o}dinger equation and calculated the eigne energy of ground states as a function of $\mu$~\cite{karsch}. Parmar et al.~\cite{parmar} used finite temperature screened potential model in Schr$\ddot{o}$dinger equation to calculate the wave function and masses of $b\bar{b}$ and $c\bar{c}$ bound states at finite temperature. Instead of using linear dependence on the distance between $q\bar{q}$ they used nonlinear dependence
    on the distance. For calculating the masses of spin singlet and triplet states, they used the zero temperature one gluon exchange term at finite temperature. Recently, a number of works have been done on calculating dissociation temperature of heavy quark ground states by using internal energy, free energy or their linear combination as a quark antiquark potential. Digel et al.~\cite{digal} used the free energy while Alberico et al.~\cite{alberico} used the internal energy in Schr{\"o}dinger equation as a quark anti-quark potential at finite temperature. They calculated temperature dependent binding energies of heavy-quark meson ground states and also predicted the dissociation temperature of these states.
  \paragraph*{}
This paper is organised as follows. In section 1 we develop non-relativistic potential model of heavy quarkonia as a function of temperature and distance. Parameters ($\alpha$, $\sigma$ and $\mu$) of our model are temperature dependent and we fit these parameters by using the lattice data of internal energy at different temperature. $\alpha(T)$, $\sigma(T)$ and $\mu(T)$ are also plotted in section 1. In section 2 we calculate spin dependent interaction terms at finite temperature by using general formulas of spin dependent terms at zero temperature. After adding these terms in our potential model, we develop complete form of potential at finite temperature. By using this complete form, we calculate the spectrum of charmonium and bottomonium at finite temperature. In this section we also calculate screening radii and root mean square radii of charmonium and bottomonium states. In section 3 we discuss results.
\section{Methodology}
\begin{figure}[H]
\centering
\includegraphics[width=14cm]{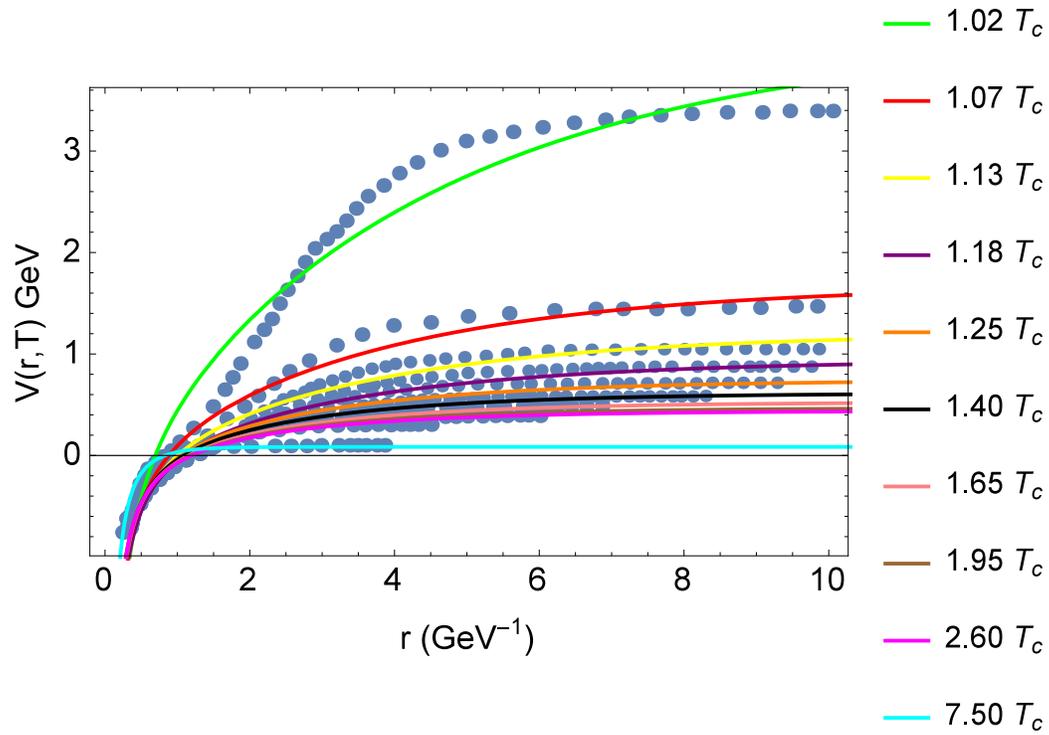}\\
  \caption{Fitted potential with lattice data from \cite{petreczky}\cite{young}. Top to bottom temperature changes from $1.02T_{c}$ to $7.50T_{c}$.  }\label{fig:fitted data}
\end{figure}

\begin{figure}[H]
\centering
\includegraphics[width=12cm]{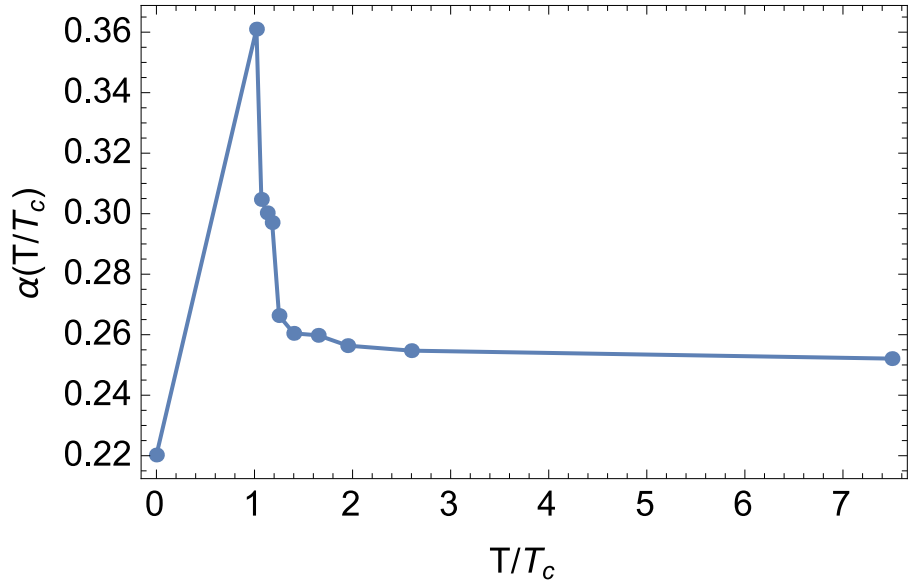}\\
  \caption{Graph of $\alpha$ as a function of temperature; dots show values of $\alpha$ from Table~\ref{tab:parameters data} and the line shows interpolated function of $\alpha(T).$ }\label{fig:alphat}
\end{figure}

\begin{figure}[H]
\centering
\includegraphics[width=12cm]{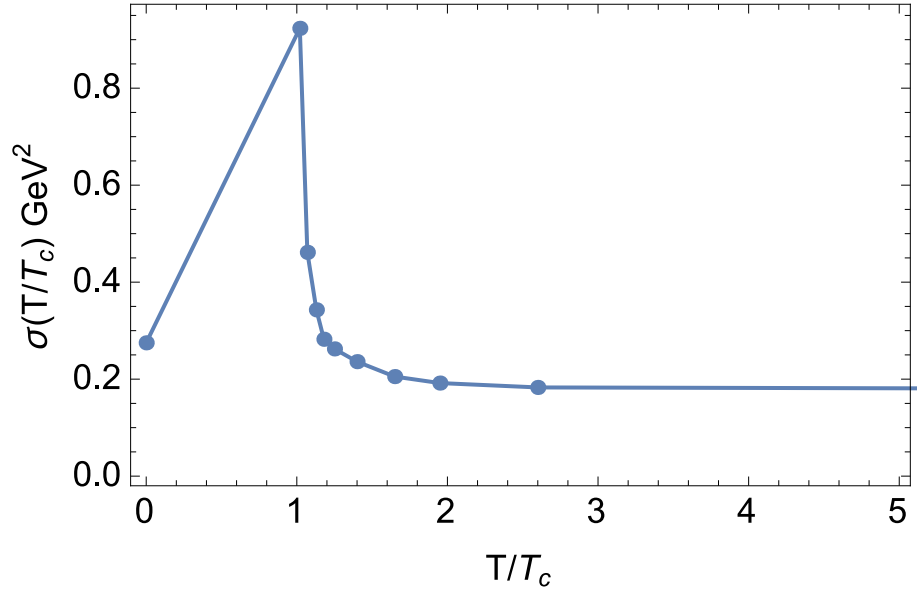}\\
  \caption{Graph of $\sigma$ as a function of temperature; dots show values of $\sigma$  from Table\ref{tab:parameters data} and the line shows interpolated function of $\sigma(T).$  }\label{fig:sigmat}
\end{figure}

\begin{figure}[H]
\centering
\includegraphics[width=12cm]{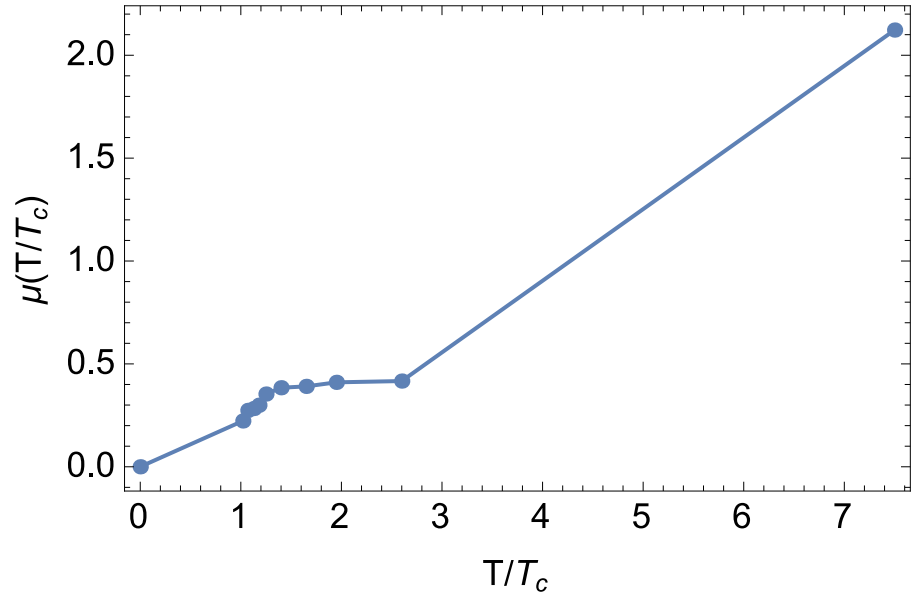}\\
  \caption{Graph of $\mu$ as a function of temperature; dots show values of $\mu$ from Table\ref{tab:parameters data} and the line shows interpolated function of $\mu(T).$}\label{fig:mut}
\end{figure}

The well-known Cornall potential used to describe the properties of heavy quarkonia very well at \emph{zero temperature} is
\begin{equation*}
    V(r, 0)  =  -\frac{4 \alpha}{3 r} + \sigma r.
\end{equation*}
Here $\alpha$ is the QCD coupling constant and $\sigma$ is the string tension for the linear confinement. As already discussed in the introduction, at finite temperature quark potential becomes screened due to the presence of color screening force. So, the modified quark anti-quark potential at finite temperature is:
\begin{equation}\label{1}
    V(r,T) = -\frac{4 \alpha(T)}{3 r} e^{-\mu(T)r} + \sigma(T)[\frac{1-e^{-\mu(T)r}}{\mu(T)}]
\end{equation}
In the above expression $\alpha(T)$, $\sigma(T)$ and $\mu(T)$ are parameters of the potential
$\alpha(T)$ is the temperature dependent coupling constant, $\sigma(T)$ is temperature dependent string tension and $\mu(T)$ is the screening mass that is inverse of screening length: $\mu(T) = \frac{1}{r_D(T)}$. At specific temperature when the screening length
 becomes less then the size of bound state, this state will dissolve. We calculated parameters of equation~\ref{1}
  by fitting
   these with lattice data of temperature dependent potential energy in quenched QCD that vary from $1.02 T_{c}$ to $7.50 T_{c}$. Fitted results are shown in Fig.~\ref{fig:fitted data} with total $\chi^2$ being $8.65772$.
   Dots represent lattice data from~\cite{petreczky},\cite{young} and lines
   show fitted potentials.
    Table~\ref{tab:parameters data} shows the screening parameters of equation (1) for different temperature. By increasing temperature, values of $\alpha$ and $\sigma$ decrease exponentially whereas $\mu$ is an increasing function of temperature, and for $T = 0$, $\mu(T) = 0$. At zero temperature values of $\alpha(0)$ and $\sigma(0)$ were calculated by fitting equation (1) with zero temperature lattice data of potential given in~\cite{juge}.
The resulting lattice fitted values of  $\alpha(0)$ and $\sigma(0)$ are
\begin{equation}\label{alpha0}
 \alpha(0) = 0.220245  ;  \sigma(0) = 0.274912\hspace{.05in} GeV^2.
 \end{equation}
 Graphs of $\alpha, \sigma$ and $\mu$ as functions of temperature are shown in Figs.~\ref{fig:alphat},\ref{fig:sigmat} and\ref{fig:mut}.

\begin{table}[H]
  \centering
 \caption{Fitted parameters of equation (1) with lattice data in temperature range 1.02 $T_{c}$ to 7.50 $T_{c}.$}\label{tab:parameters data}
  \begin{tabular}{|c|c|c|c|}
    \hline
     Temperature & $\alpha(T)$ & $\sigma(T)$ & $\mu(T)$   \\[1ex]
          $T_{c}$ &            &    $GeV^2$ & GeV\\[1ex]
    \hline
    \hline
     1.02 & 0.361006 & 0.923378 & 0.222732 \\[1ex]
     1.07 & 0.304707 & 0.461444 & 0.274135 \\[1ex]
     1.13 & 0.300275 & 0.342902 & 0.28331  \\[1ex]
     1.18 & 0.297069 & 0.282231 & 0.299379 \\[1ex]
     1.25 & 0.266321 & 0.262375 & 0.353513 \\[1ex]
     1.40 & 0.26045  & 0.236071 & 0.38413  \\[1ex]
     1.65 & 0.25979  & 0.205274 & 0.390749 \\[1ex]
     1.95 & 0.256385 & 0.191909 & 0.410369 \\[1ex]
     2.60 & 0.254726 & 0.182913 & 0.416623 \\[1ex]
     7.50 & 0.252093 & 0.17934  & 2.12253  \\[1ex]
     \hline

  \end{tabular}
\end{table}

\section{Spin dependent finite temperature potential terms}
In order to avoid spin degenerate terms we added spin-dependent effects at finite temperature in our potential model that controlled the fine and hyperfine splitting in the spectrum. Spin dependent corrections contributed three types of interaction terms in the potential, same as those zero temperature potential model\cite{lucha} \cite{voloshin}.
\begin{equation}
V_{spin}(r, T) = V_{LS} (r, T) \textbf{L.S} + V_T(r, T) \textbf{T} + V_{SS}(r, T) \textbf{S}_1.\textbf{S}_2
\end{equation}
In above equation $V_{LS}$ is spin-orbit term, $V_T$ is tensor term and $V_{SS}$ is spin term which is used to differentiate between spin singlet and triplet states.
The values of $\textbf{L.S} = \frac{[J(J+1)-L(L+1)-S(S+1)]}{2}$ , $\textbf{S}_1 .\textbf{S}_2 = [\frac{S(S+1)}{2} -\frac{3}{4}]$ and non vanishing diagonal matrix element of $\textbf{T}$ for $l\neq 0$ and $s = 1$ are:
\begin{equation*}
    \langle^3 L_J|T|^3 L_J\rangle =
\begin{cases}
    -\frac{1}{6(2L+3)}, &\text{J=L+1}\\
       +\frac{1}{6}, &\text{J=L}\\
      -\frac{L+1}{6(2L-1)}, &\text{J=L-1.}
\end{cases}
\end{equation*}
With dividing equation (1) in vector $V_v(r, T)$ and scalar $V_s(r, T)$ parts of potential, the general forms of spin-spin, spin-orbit and tensor terms \cite{lucha} are written below:
\begin{equation}
    V_{LS} = \frac{1}{2 m_1^2 r}[3\frac{d}{dr}V_v(r,T) -\frac{d}{dr} V_s(r,T)]
\end{equation}
and
\begin{equation}
    V_{T} = \frac{1}{m_1^2}[\frac{1}{r}\frac{d}{dr}V_v(r,T) - \frac{d^{2}}{dr^{2}}V_v(r,T)]
\end{equation}
\begin{equation}
  V_{SS} = \frac{2}{3 m_1^2}\Delta V_v(r, T).
\end{equation}
In above three equations $m_1$ is the mass of charm or bottom quark, $V_v(r,T) = -\frac{4 \alpha(T)}{3 r} e^{-\mu(T)r} $ and $V_s(r,T) = \sigma(T)[\frac{1-e^{-\mu(T)r}}{\mu(T)}]$. $\Delta = \nabla^2$ is Laplacian. We calculated the derivative of vector and scalar part of equation (1) potential, put in equation (4), (5), (6) and used in equation (3) to calculate the spin dependent part of the potential that is given below:
\begin{equation}
  V_{LS}(r, T) = \frac{1}{2 m_1^2 r^3}[ e^{-\mu(T)r}  (4\alpha(T)+ 4 \alpha(T)\mu(T) r ) -\sigma(T) r^2]
\end{equation}
\begin{equation}
V_T (r, T) = \frac{4 \alpha(T) e^{-\mu(T) r }}{3 m_1^2 r^3 }[ 3 + 3 \mu(T) r + \mu(T)^2 r^2]
\end{equation}
\begin{equation}
V_{SS}(r, T) = \frac{-8\alpha(T) \mu(T)^2}{9 m_1^2 r} e^{-\mu(T)r}.
\end{equation}
So, complete form of finite temperature potential $V(r, T)$ between $q\bar{q}$ is given by:
\begin{equation}\label{finite}
\begin{split}
    V(r, T) = -\frac{4 \alpha(T)}{3 r} e^{-\mu(T)r} + \sigma(T)[\frac{1-e^{-\mu(T)r}}{\mu(T)}] +  & \frac{-8\alpha(T) \mu(T)^2}{9 m_1^2 r}\textbf{S}_1.\textbf{S}_2\\
     & + \frac{1}{2 m_1^2 r^3}[ e^{-\mu(T)r}  (4\alpha(T)+ 4 \alpha(T)\mu(T) r ) -\sigma(T) r^2]\textbf{L.S}\\
     & +\frac{4 \alpha(T) e^{-\mu(T) r }}{3 m_1^2 r^3 }[ 3 + 3 \mu(T) r + \mu(T)^2 r^2]\textbf{T}.
\end{split}
\end{equation}

For calculating the temperature dependent spectrum of heavy quark anti-quark bound state we used the radial Schr{\"o}dinger equation with potential given in equation~\ref{finite}. Phenomenologically, when we calculated the spectrum of $b\bar{b}$ and $c\bar{c}$ unknown parameters in zero temperature potential, $\alpha_s (0)$, $\sigma(0)$, $m_b$ and $m_c$ were fixed by using the masses of known states of charmonia and botomonia. At zero temperature fitted values of these parameters were: for $\emph{c}\bar{c}$ $\alpha(0) = 0.5461$, $\sigma(0) = 0.1425 GeV^2$ and $m_c = 1.4794 GeV$ \cite{sultan}. In case of $\emph{b}\bar{b}$ $\alpha(0) = 0.36$, $\sigma(0) = 0.1340 GeV^2$ $m_b = 4.825 GeV$. $\alpha(0)$ and $\sigma(0)$  can also be fitted by using lattice data of potential energy at zero temperature \cite{juge}. But the calculated  masses of charmonia and bottomonia by using lattice fitted values of parameters did not agree with experiment.
 At finite temperature $\alpha(T)$ and $\sigma(T)$ were fitted by using lattice data of potential energy at finite temperature as in Table.\ref{tab:parameters data}. But the calculated spectrum with lattice fitted parameters did not agree with experimental data that create problem at finite temperature.
  To solve this problem at finite temperature we calculated the difference between lattice fitted parameters and phenomenologically fitted parameters at zero temperature and multiplied these difference factors
   with $\alpha(T)$ and $\sigma(T)$ at finite temperature. For charmonium multiplicative factor
    for $\alpha(T)$ is $2.47951$ and $\sigma(T)$ is multiplied with $0.518348$, while in case of bottomonium $\alpha(T)$ is multiplied with $1.63454$ and $\sigma(T)$ with $0.487429$.
    For calculating temperature dependent spectrum of charmonium and bottomoinum meson we used non-relativistic Sch{\"o}dinger equation \ref{sch} and V(r,T) taken from \ref{finite}
\begin{equation} \label{sch}
  U^{\prime\prime} (r) + 2 m (E - V(r, T) - \frac{<L^{2}_{q\overline{q}}>}{2 m r^{2}})U(r) = 0.
\end{equation}
In equation~\ref{sch} $E$ is the total energy of the bound state, $U(r) = r R(r)$ where $R(r)$ is the radial wave function, $m = \frac{m_1 m_2}{m_1 + m_2}$ is the reduced mass of quark antiquark system where $m_1$ and $m_2$ are the masses of quark and antiquark respectively. $<L^{2}_{q\overline{q}}> = l(l+1)$ is the relative angular momentum of quark antiquark system. We used shooting method for calculating the non-trivial solutions of equation \ref{sch} for certain discrete values of $E$. For masses of $\emph{b}\bar{b}$ and $\emph{c}\bar{c}$ states we added their constituents masses (i.e. $2m_b$ or $2m_c$) in the total energy. It is to be noted that at short distance $\frac{1}{r^3}$ term in the potential dominates and thus
for $S = 1$, $J = L$ and $J = L-1$ $\textbf{L.S} + 2T$ becomes negative making the potential strongly attractive.
This resulted in an unstable wave function at $r \rightarrow 0$. To overcome this problem we first solved Schr{\"o}dinger equation without spin-orbit coupling term and calculated the masses. The spin-orbit terms were
 incorporated through the leading order perturbative correction and added in the masses of the bound state~\cite{akram}. The resulting masses for charmonium meson are in Table~\ref{tab:cc} and for bottominum meson in Table~\ref{tab:bb}. We also calculate root mean square radii and screening radii for charmonium and bottomonium states.
\section{Results}
We used lattice data of static quark anti-quark internal energy to fit a temperature dependent potential model. At finite temperature entropy affects the quark anti-quark interaction and in internal energy data
entropy contribution is also included.
\begin{equation*}
U(r, T) = F(r, T) + T S(r, T)
\end{equation*}
Here $U(r, T)$ is the internal energy, $F(r, T)$ is the free energy and $S(r, T)$ is the entropy.\newline
Fitted graphs of our potential model
for different temperature are shown in Fig.~\ref{fig:fitted data}. When $T < T_c$ and at small distance our finite temperature potential coincide with the  $T = 0$ potential, and at large distance by increasing temperature potential is also increased.
When temperature of the medium is increased by critical temperature then potential is decreased by increasing temperature.
 Our potential model agrees with Fig. 4
 of~\cite{zantow} and Fig. 3 of \cite{kaczmarek}. Graph of $\alpha(T)$ and $\sigma(T)$
 is shown in Fig.~\ref{fig:alphat} and \ref{fig:sigmat}. In the limit $T<T_c$, values of $\alpha$ and $\sigma$
 increase with temperature and after critical temperature both
 decreases with temperature.\newline
We also studied the effect of medium on charmonium and bottomoium spectrum by solving Schrs$\ddot{o}$dinger equation with finite temperature potential model.
 We calculated critical temperature of the states
 by checking the spectrum of the states. At particular temperature spectrum of a state becomes unstable. This temperature is called the critical temperature of the state. We can see in the Tables.~\ref{tab:cc} and \ref{tab:bb} that ground states of charmonia and bottomonia are tightly bound and survive in hot medium. 
 We also calculate masses and root mean square radii at critical temperature.
 Corresponding to critical temperature of a particular state we calculate its screening radii by using $r_D = 1/\mu(T_c)$. Results are shown in Table.~\ref{tab:cc} and \ref{tab:bb}. At critical temperature screening radii $r_D$ are less then root mean square radii $r_o$
 . $J/\psi$ and $\eta_c$ survive in hot medium till the temperature reaches at 2.78$T_c$
  and the survival probability of $\Upsilon$ and $\eta_b$ is 3.5 to 4.00 $T_c$~\cite{2016}.
  From the tables we can see that spin singlet and spin triplet states of $c\bar{c}$ and $b\bar{b}$ dissolve on that same temperature.
   By changing the principle quantum number $"n"$
   and the orbital quantum number $"l"$ critical temperatures of the states also change. The states with different angular quantum number "$j"$ also dissolve at the same temperature. So critical temperature of the states depends upon $"n"$ and $"l"$ and not on $"s"$ and $"j"$.
\begin{table}[H]
  \centering
  \caption{Calculated spectrum of \emph{c}$\bar{c}$ meson at finite temperature.}\label{tab:cc}
 \begin{tabular}{|c|c|c|c|c|c|>{\centering\arraybackslash}m{2cm}|>{\centering\arraybackslash}m{3cm}|c|c|}
    \hline
     n & Meson & L & S & J & $J^{PC}$ & Critical Temperature & Mass & rms value at $T_c$ & screening radius $r_D$\\[1ex]
    \hline
    & & & & & & GeV & GeV & $GeV^{-1}$ & $GeV^{-1}$\\[1ex]
    \hline\hline
      1S& $\eta_c$ (1${}^1{S}_0$) & 0 & 0 & 0 & $0^{-+}$ & 2.78 & 3.4233 & 11.9758& 2.08642 \\[1ex]
        & $J/\psi$ (1${}^3{S}_1$) & 0 & 1 & 1 & $1^{--}$ & 2.78 & 3.42138 & 12.0338 & 2.08642\\[1ex]
    \hline
      2S& $\eta_c$ (2${}^1{S}_0$) & 0 & 0 & 0 & $0^{-+}$ & 1.18 & 3.51905 & 12.6636 & 3.34025 \\[1ex]
        & $J/\psi$ (2${}^3{S}_1$) & 0 & 1 & 1 & $1^{--}$ & 1.18 & 3.51848 & 12.7256 & 3.34025\\[1ex]
    \hline
      3S& $\eta_c$ (3${}^1{S}_0$) & 0 & 0 & 0 & $0^{-+}$ & 1.08 & 3.99659 & 12.5784 & 3.6276 \\[1ex]
        & $J/\psi$ (3${}^3{S}_1$) & 0 & 1 & 1 & $1^{--}$ & 1.08 & 3.99586 & 12.563 & 3.6276\\[1ex]
    \hline
      4S& $\eta_c$ (4${}^1{S}_0$) & 0 & 0 & 0 & $0^{-+}$ & 1.05 & 4.51806 & 13.1879 & 3.94363\\[1ex]
        & $J/\psi$ (4${}^3{S}_1$) & 0 & 1 & 1 & $1^{--}$ & 1.05 & 4.51739 & 13.1941 & 3.94363\\[1ex]
       \hline
      1P& $\emph{h}_c$ (1${}^1{P}_1$) & 1 & 0 & 1 & $1^{+-}$ & 1.18 & 3.60595 & 12.099  & 3.34025 \\[1ex]
        & $\chi_0$ (1${}^3{P}_0$)     & 1 & 1 & 0 & $0^{++}$ & 1.18 & 3.57631 & 12.1248 & 3.34025 \\[1ex]
        & $\chi_1$ (1${}^3{P}_1$)     & 1 & 1 & 1 & $1^{++}$ & 1.18 & 3.59179 & 12.1248 & 3.34025\\[1ex]
        & $\chi_2$ (1${}^3{P}_2$)     & 1 & 1 & 2 & $2^{++}$ & 1.18 & 3.61946 & 12.1248 & 3.34025\\[1ex]
    \hline
      2P& $\emph{h}_c$ (2${}^1{P}_1$) & 1 & 0 & 1 & $1^{+-}$ & 1.08 & 4.12356 & 12.088  & 3.6276  \\[1ex]
        & $\chi_0$ (2${}^3{P}_0$)     & 1 & 1 & 0 & $0^{++}$ & 1.08 & 4.07012 & 12.0821 & 3.6276\\[1ex]
        & $\chi_1$ (2${}^3{P}_1$)     & 1 & 1 & 1 & $1^{++}$ & 1.08 & 4.09818 & 12.0821 & 3.6276\\[1ex]
        & $\chi_2$ (2${}^3{P}_2$)     & 1 & 1 & 2 & $2^{++}$ & 1.08 & 4.14844 & 12.0821 & 3.6276\\[1ex]
    \hline
      3P& $\emph{h}_c$ (3${}^1{P}_1$) & 1 & 0 & 1 & $1^{+-}$ & 1.05 & 4.49963 & 12.9823 & 3.94363\\[1ex]
        & $\chi_0$ (3${}^3{P}_0$)     & 1 & 1 & 0 & $0^{++}$ & 1.05 & 4.44281 & 12.9767 & 3.94363\\[1ex]
        & $\chi_1$ (3${}^3{P}_1$)     & 1 & 1 & 1 & $1^{++}$ & 1.05 & 4.47275 & 12.9767 & 3.94363 \\[1ex]
        & $\chi_2$ (3${}^3{P}_2$)     & 1 & 1 & 2 & $2^{++}$ & 1.05 & 4.5263  & 12.9767 & 3.94363\\[1ex]
    \hline
      4P& $\emph{h}_c$ (4${}^1{P}_1$) & 1 & 0 & 1 & $1^{+-}$ & 1.02 & 5.20158 & 12.8187 & 4.4897 \\[1ex]
        & $\chi_0$ (4${}^3{P}_0$)     & 1 & 1 & 0 & $0^{++}$ & 1.02 & 5.24053 & 13.3454 & 4.4897\\[1ex]
        & $\chi_1$ (4${}^3{P}_1$)     & 1 & 1 & 1 & $1^{++}$ & 1.02 & 5.28607 & 13.3454 & 4.4897\\[1ex]
        & $\chi_2$ (4${}^3{P}_2$)     & 1 & 1 & 2 & $2^{++}$ & 1.02 & 5.36757 & 13.3454 & 4.4897\\[1ex]
    \hline

  \end{tabular}
\end{table}
\begin{table}[H]
  \centering
  \caption{Calculated spectrum of \emph{c}$\bar{c}$ meson at finite temperature (continued).}
  \begin{tabular}{|c|c|c|c|c|c|>{\centering\arraybackslash}m{2cm}|>{\centering\arraybackslash}m{3cm}|c|c|}
    \hline
     n & Meson & L & S & J & $J^{PC}$ & Critical Temperature & Mass & rms value at $T_c$ & screening radius $r_D$\\[1ex]
    \hline
    & & & & & & GeV & GeV & $GeV^{-1}$ & $GeV^{-1}$\\[1ex]
    \hline\hline
      1D& $\eta_{c2}$ (1${}^1{D}_2$)  & 2 & 0 & 2 & $2^{-+}$ & 1.08 & 4.0991  & 11.689  & 3.6276 \\[1ex]
        & $\psi$ (1${}^3{D}_1$)       & 2 & 1 & 1 & $1^{--}$ & 1.08 & 4.08603 & 11.6887 & 3.6276 \\[1ex]
        & $\psi_2$ (1${}^3{D}_2$)     & 2 & 1 & 2 & $2^{--}$ & 1.08 & 4.09474 & 11.6887 & 3.6276\\[1ex]
        & $\psi_3$ (1${}^3{D}_3$)     & 2 & 1 & 3 & $3^{--}$ & 1.08 & 4.10701 & 11.6887 & 3.6276 \\[1ex]
    \hline

      2D& $\eta_{c2}$ (2${}^1{D}_2$)  & 2 & 0 & 2 & $2^{-+}$ & 1.05 & 4.62788 & 12.4265 & 3.94363 \\[1ex]
        & $\psi$ (2${}^3{D}_1$)       & 2 & 1 & 1 & $1^{--}$ & 1.05 & 4.60899 & 12.4219 & 3.94363\\[1ex]
        & $\psi_2$ (2${}^3{D}_2$)     & 2 & 1 & 2 & $2^{--}$ & 1.05 & 4.62168 & 12.4219 & 3.94363\\[1ex]
        & $\psi_3$ (2${}^3{D}_3$)     & 2 & 1 & 3 & $3^{--}$ & 1.05 & 4.63955 & 12.4219 & 3.94363\\[1ex]
    \hline
      3D& $\eta_{c2}$ (3${}^1{D}_2$)  & 2 & 0 & 2 & $2^{-+}$ & 1.04 & 4.65541 & 13.5951 & 4.11027 \\[1ex]
        & $\psi$ (3${}^3{D}_1$)       & 2 & 1 & 1 & $1^{--}$ & 1.04 & 4.64297 & 13.5915 & 4.11027\\[1ex]
        & $\psi_2$ (3${}^3{D}_2$)     & 2 & 1 & 2 & $2^{--}$ & 1.04 & 4.65135 & 13.5915 & 4.11027\\[1ex]
        & $\psi_3$ (3${}^3{D}_3$)     & 2 & 1 & 3 & $3^{--}$ & 1.04 & 4.66307 & 13.5915 & 4.11027\\[1ex]
    \hline
      4D& $\eta_{c2}$ (4${}^1{D}_2$)  & 2 & 0 & 2 & $2^{-+}$ & 0.96 & 5.40349 & 13.1152 & 4.77031 \\[1ex]
        & $\psi$ (4${}^3{D}_1$)       & 2 & 1 & 1 & $1^{--}$ & 0.96 & 5.37742 & 13.123  & 4.77031 \\[1ex]
        & $\psi_2$ (4${}^3{D}_2$)     & 2 & 1 & 2 & $2^{--}$ & 0.96 & 5.39517 & 13.123  & 4.77031\\[1ex]
        & $\psi_3$ (4${}^3{D}_3$)     & 2 & 1 & 3 & $3^{--}$ & 0.96 & 5.42015 & 13.123  & 4.77031\\[1ex]
    \hline
  \end{tabular}
\end{table}

\begin{table}[H]
  \centering
  \caption{Calculated spectrum of \emph{b}$\bar{b}$ meson at finite temperature.}\label{tab:bb}
   \begin{tabular}{|c|c|c|c|c|c|>{\centering\arraybackslash}m{2cm}|>{\centering\arraybackslash}m{3cm}|c|c|}
    \hline
     n & Meson & L & S & J & $J^{PC}$ & Critical Temperature & Mass & rms value at $T_c$ & screening radius $r_D$\\[1ex]
    \hline
    & & & & & & GeV & GeV & $GeV^{-1}$ & $GeV^{-1}$\\[1ex]
    \hline\hline
      1S& $\eta_b$ (1${}^1{S}_0$) & 0 & 0 & 0 & $0^{-+}$   & 4.00 & 9.87288 & 12.0228 &1.10616 \\[1ex]
        & $\Upsilon$ (1${}^3{S}_1$) & 0 & 1 & 1 & $1^{--}$ & 4.00 & 9.92965 & 11.0974 & 1.10616\\[1ex]
    \hline
      2S& $\eta_b$ (2${}^1{S}_0$) & 0 & 0 & 0 & $0^{-+}$   & 2.69 & 9.86967 & 12.9578 & 2.23236\\[1ex]
        & $\Upsilon$ (2${}^3{S}_1$) & 0 & 1 & 1 & $1^{--}$ & 2.69 & 9.86963 & 12.9707 & 2.23236\\[1ex]
    \hline
      3S& $\eta_b$ (3${}^1{S}_0$) & 0 & 0 & 0 & $0^{-+}$ & 1.23 & 10.0915 & 12.1396 & 2.95818\\[1ex]
        & $\Upsilon$ (3${}^3{S}_1$) & 0 & 1 & 1 & $1^{--}$ & 1.23 & 10.0914 & 12.14 & 2.95818\\[1ex]
    \hline
      4S& $\eta_b$ (4${}^1{S}_0$) & 0 & 0 & 0 & $0^{-+}$   & 1.12 & 10.3351 & 12.5704 & 3.54886\\[1ex]
        & $\Upsilon$ (4${}^3{S}_1$) & 0 & 1 & 1 & $1^{--}$ & 1.12 & 10.335  & 12.5686 & 3.54886\\[1ex]
       \hline
      1P& $\emph{h}_b$ (1${}^1{P}_1$) & 1 & 0 & 1 & $1^{+-}$ & 2.69 & 9.87014 & 12.6705 & 2.23236\\[1ex]
        & $\chi_0$ (1${}^3{P}_0$)     & 1 & 1 & 0 & $0^{++}$ & 2.69 & 9.86864 & 12.6788 &2.23236\\[1ex]
        & $\chi_1$ (1${}^3{P}_1$)     & 1 & 1 & 1 & $1^{++}$ & 2.69 & 9.86942 & 12.6788 &2.23236 \\[1ex]
        & $\chi_2$ (1${}^3{P}_2$)     & 1 & 1 & 2 & $2^{++}$ & 2.69 & 9.87081 & 12.6788 &2.23236\\[1ex]
    \hline
      2P& $\emph{h}_b$ (2${}^1{P}_1$) & 1 & 0 & 1 & $1^{+-}$ & 1.23 & 10.0892 & 12.0998 &2.95818\\[1ex]
        & $\chi_0$ (2${}^3{P}_0$)     & 1 & 1 & 0 & $0^{++}$ & 1.23 & 10.0852 & 12.1027 &2.95818\\[1ex]
        & $\chi_1$ (2${}^3{P}_1$)     & 1 & 1 & 1 & $1^{++}$ & 1.23 & 10.0873 & 12.1027 &2.95818\\[1ex]
        & $\chi_2$ (2${}^3{P}_2$)     & 1 & 1 & 2 & $2^{++}$ & 1.23 & 10.0911 & 12.1027 &2.95818\\[1ex]
    \hline
      3P& $\emph{h}_b$ (3${}^1{P}_1$) & 1 & 0 & 1 & $1^{+-}$ & 1.11 & 10.4685  & 12.4192 & 3.56822\\[1ex]
        & $\chi_0$ (3${}^3{P}_0$)     & 1 & 1 & 0 & $0^{++}$ & 1.11 & 10.4598  & 12.4181 & 3.56822\\[1ex]
        & $\chi_1$ (3${}^3{P}_1$)     & 1 & 1 & 1 & $1^{++}$ & 1.11 & 10.4644 & 12.4181  & 3.56822\\[1ex]
        & $\chi_2$ (3${}^3{P}_2$)     & 1 & 1 & 2 & $2^{++}$ & 1.11 & 10.4727 & 12.4181  & 3.56822\\[1ex]
    \hline
      4P& $\emph{h}_b$ (4${}^1{P}_1$) & 1 & 0 & 1 & $1^{+-}$ & 1.07 & 10.4987 & 12.6889 & 3.64784\\[1ex]
        & $\chi_0$ (4${}^3{P}_0$)     & 1 & 1 & 0 & $0^{++}$ & 1.07 & 10.493  & 12.6869 & 3.64784\\[1ex]
        & $\chi_1$ (4${}^3{P}_1$)     & 1 & 1 & 1 & $1^{++}$ & 1.07 & 10.496  & 12.6869 & 3.64784\\[1ex]
        & $\chi_2$ (4${}^3{P}_2$)     & 1 & 1 & 2 & $2^{++}$ & 1.07 & 10.5015 & 12.6869 & 3.64784\\[1ex]
    \hline

  \end{tabular}
\end{table}
\begin{table}[H]
  \centering
  \caption{Calculated spectrum of \emph{b}$\bar{b}$ meson at finite temperature (continued).}
  \begin{tabular}{|c|c|c|c|c|c|>{\centering\arraybackslash}m{2cm}|>{\centering\arraybackslash}m{3cm}|c|c|}
    \hline
     n & Meson & L & S & J & $J^{PC}$ & Critical Temperature & Mass & rms value at $T_c$ & screening radius $r_D$\\[1ex]
    \hline
    & & & & & & GeV & GeV & $GeV^{-1}$ & $GeV^{-1}$\\[1ex]
    \hline\hline
      1D& $\eta_{b2}$ (1${}^1{D}_2$)      & 2 & 0 & 2 & $2^{-+}$ & 1.23 & 10.0867 &12.2507 &2.95818\\[1ex]
        & $\Upsilon$ (1${}^3{D}_1$)       & 2 & 1 & 1 & $1^{--}$ & 1.23 & 10.0858 &12.2542 & 2.95818\\[1ex]
        & $\Upsilon_2$ (1${}^3{D}_2$)     & 2 & 1 & 2 & $2^{--}$ & 1.23 & 10.0864 &12.2542 & 2.95818 \\[1ex]
        & $\Upsilon_3$ (1${}^3{D}_3$)     & 2 & 1 & 3 & $3^{--}$ & 1.23 & 10.0873 &12.2542 & 2.95818\\[1ex]
    \hline

      2D& $\eta_{b2}$ (2${}^1{D}_2$)      & 2 & 0 & 2 & $2^{-+}$ & 1.10 & 10.4423 & 12.386 & 3.5878\\[1ex]
        & $\Upsilon$ (2${}^3{D}_1$)       & 2 & 1 & 1 & $1^{--}$ & 1.10 & 10.4402 & 12.3875& 3.5878\\[1ex]
        & $\Upsilon_2$ (2${}^3{D}_2$)     & 2 & 1 & 2 & $2^{--}$ & 1.10 & 10.4417 & 12.3875& 3.5878\\[1ex]
        & $\Upsilon_3$ (2${}^3{D}_3$)     & 2 & 1 & 3 & $3^{--}$ & 1.10 & 10.4437 & 12.3875& 3.5878\\[1ex]
    \hline
      3D& $\eta_{b2}$ (3${}^1{D}_2$)      & 2 & 0 & 2 & $2^{-+}$ & 1.07 & 10.4961 & 12.7232 & 3.64784\\[1ex]
        & $\Upsilon$ (3${}^3{D}_1$)       & 2 & 1 & 1 & $1^{--}$ & 1.07 & 10.4946 & 12.7284 & 3.64784\\[1ex]
        & $\Upsilon_2$ (3${}^3{D}_2$)     & 2 & 1 & 2 & $2^{--}$ & 1.07 & 10.4956 & 12.7284 & 3.64784 \\[1ex]
        & $\Upsilon3$ (3${}^3{D}_3$)      & 2 & 1 & 3 & $3^{--}$ & 1.07 & 10.4972 & 12.7284 & 3.64784\\[1ex]
    \hline
      4D& $\eta_{b2}$ (4${}^1{D}_2$)      & 2 & 0 & 2 & $2^{-+}$ & 1.055 & 10.8444 & 13.2544 & 3.86527 \\[1ex]
        & $\Upsilon$ (4${}^3{D}_1$)       & 2 & 1 & 1 & $1^{--}$ & 1.055 & 10.8418 & 13.2529 & 3.86527\\[1ex]
        & $\Upsilon_2$ (4${}^3{D}_2$)     & 2 & 1 & 2 & $2^{--}$ & 1.055 & 10.8436 & 13.2529 & 3.86527\\[1ex]
        & $\Upsilon_3$ (4${}^3{D}_3$)     & 2 & 1 & 3 & $3^{--}$ & 1.055 & 10.8461 & 13.2529 & 3.86527\\[1ex]
    \hline
  \end{tabular}
\end{table}

\end{document}